No substitute for real data: phylogenies from birth-death polytomy resolvers should not be used for many downstream comparative analyses

Daniel L. Rabosky

Department of Ecology and Evolutionary Biology and Museum of Zoology

University of Michigan, Ann Arbor, MI 48103

Email: drabosky@umich.edu

Article type: Brief communication

Running head: Polytomy resolvers can bias comparative analyses

Data Archival Location:

Abstract

The statistical estimation of phylogenies is always associated with uncertainty, and accommodating this uncertainty is an important component of modern phylogenetic comparative analysis. The birth-death polytomy resolver is a method of accounting for phylogenetic uncertainty that places missing (unsampled) taxa onto phylogenetic trees, using taxonomic information alone. Recent studies of birds and mammals have used this approach to generate pseudo-posterior distributions of phylogenetic trees that are complete at the species level, even in the absence of genetic data for many species. Many researchers have used these distributions of phylogenies for downstream evolutionary analyses that involve inferences on phenotypic evolution, geography, and community assembly. I demonstrate that the use of phylogenies constructed in this fashion is inappropriate for many questions involving traits. Because species are placed on trees at random with respect to trait values, the birth-death polytomy resolver breaks down natural patterns of trait phylogenetic structure. Inferences based on these trees are predictably and often drastically biased in a direction that depends on the underlying (true) pattern of phylogenetic structure in traits. I illustrate the severity of the phenomenon for both continuous and discrete traits using examples from a global bird phylogeny.

Introduction

Phylogenies have become essential for addressing many fundamental questions in ecology and evolutionary biology. Many features of organisms, from morphological traits to ecological community membership, bear the imprint of phylogenetic history. As Felsentein (1985) noted, "Phylogenies are fundamental to comparative biology: there is no doing it without taking them into account." The rise of phylogenetic comparative methods (Felsenstein 1985; Grafen 1989; Harvey and Pagel 1991) and convenient statistical packages that implement them (Paradis et al. 2004; Revell 2012; Maddison 2015) has made it easy for researchers to incorporate a phylogenetic perspective into evolutionary research. In parallel with the availability of phylogenetic methods, the availability of phylogenetic data has increased dramatically. The rise of modern statistical phylogenetics (Felsenstein 2004) has changed our conception of what a phylogeny represents: as statistical estimates, phylogenies include estimation error associated with both the pattern of branching (topology) as well as branch lengths. Through explicit simulation of posterior distributions of phylogenetic trees, Bayesian phylogenetics (Huelsenbeck et al. 2001) has provided us with considerable insight into the magnitude of uncertainty in phylogenetic tree estimates.

Many researchers recognize the importance of accommodating phylogenetic uncertainty in macroevolutionary and comparative analysis. Hereafter, I refer to any derived uses of phylogenies — for inference on trait evolution, comparative biology, phylogenetic community structure, and so on — as a "downstream comparative analysis." Because phylogenies are estimated with error, it is important to assess whether the results of comparative analyses are robust to the uncertainty that is inherent to the phylogenies that underlie them. A common practice involves performing macroevolutionary or comparative analyses across a posterior distribution of phylogenetic trees to assess the effects of phylogenetic uncertainty on inference, or to derive quasi-posterior distributions of macroevolutionary rate parameters.

Despite the necessity of a phylogenetic perspective in evolutionary biology, we are still far from complete phylogenetic knowledge for any major (> 5,000 species) group of organisms (Reddy 2014). Among clades above this size, we currently have DNA-sequence based phylogenies for approximately 67% of extant bird species (Jetz et al. 2012), 55% of mammals (Bininda-Emonds et al. 2007), 27% of fishes (Rabosky et al. 2013), and 45% of squamate reptiles (Pyron et al. 2013). A recent DNA sequence-based phylogeny of land plants (Zanne et al. 2014) is approximately 13% complete at the species level. Achieving a DNA sequence based phylogeny for a major clade of organisms that is more than 95% complete at the species level will be a significant milestone in our understanding of biological diversity. Unsampled taxa are a phylogenetically non-random subset of the tree of life, and they are disproportionately found in tropical clades (Thomson and Schaeffer 2010; Reddy 2014). In general, given the ease with which DNA sequence data can now be obtained, it is reasonable to assume that the availability of the tissue samples themselves is often the limiting factor to completing these trees.  Samples for many taxa currently lacking representation on Genbank will be nontrivial to obtain: the taxa may be rare in their native range, or their collection may require expensive and logistically-demanding expeditionary fieldwork to largely inaccessible regions.

Recently, it has become possible to impute the position of missing (unsampled) taxa into phylogenetic trees using information from taxonomic hierarchies alone, enabling researchers to generate phylogenies that are effectively complete at the species level despite the lack of genetic data for many taxa. The birth-death polytomy resolvers (BDPR) of Kuhn et al. (Kuhn et al. 2011) and Thomas et al. (2013) can be used to generate distributions of fully resolved phylogenies, where the positions of missing taxa are constrained by taxonomy and represent stochastic realizations of a constant-rate birth-death process. For example, consider a genus where 60% of known taxonomic diversity has been included in a DNA sequence-based phylogenetic analysis. Using the approach of Kuhn et al. (2011), it is possible to generate a distribution of placements for the unsampled 40% of taxa, conditional on the assumption that the genus is indeed monophyletic. Kuhn et al. (2011) demonstrated that the BDPR approach generates conservative placements of missing taxa with respect to divergence times and diversification rate estimation, providing a significant advance for diversification studies. Because the missing taxa are placed on the tree under what is effectively a null hypothesis (constant-rate diversification), inferences about diversification rates made with such trees are generally expected to be conservative (Kuhn et al. 2011). This approach has been used to study diversification dynamics across large phylogenies where the positions of a large number of taxa (33% for birds) have been simulated (Jetz et al. 2012). A new software tool (PASTIS) enables researchers to easily generate similar distributions of phylogenies for any group of organisms, provided that some set of taxonomic constraints are available for unsampled taxa (Thomas et al. 2013). The Kuhn et al. (2011) algorithm generates taxon placements that differ slightly from PASTIS, but both approaches are expected to generate distributions of missing taxon placements that are conservative with respect to diversification analyses.

Pseudoposterior distributions of phylogenetic trees for both birds and mammals have been constructed using the BDPR method and have been made conveniently available for the research community (e.g., http://birdtree.org). These phylogenies include nearly all recognized taxonomic diversity of each group and represent a significant achievement in our understanding the phylogenetic structure of vertebrate diversity. Many researchers have now used these trees in downstream comparative analyses, with the assumption that the BDPR phylogenies are a conservative estimate of phylogenetic uncertainty (e.g., Rolland et al. 2014; Price et al. 2012; Feeney et al. 2013; Gianuca et al. 2014; Healy et al. 2014). These uses have included phylogenetic community structure, phylogenetic generalized least-squares regression, and analysis of trait-dependent speciation-extinction rates. Rubolini et al. (2015) suggested that the pseudoposterior distribution of avian phylogenies from Jetz et al. (2012) should become the *de facto* standard for accommodating uncertainty in downstream comparative analyses for the foreseeable future.

In this article, I demonstrate that although these methods are useful for estimates of diversification rates, the use of BDPR phylogenies is generally not appropriate for uses that involve trait data.This includes analyses that involve morphology, ecology, geography, community membership, and the relationship between any such traits and species diversification. This problem occurs because BDPR phylogenies involve conservative placements of unsampled taxa with respect to a birth-death process, but not with respect to species trait values. For traits that are phylogenetically conserved or overdispersed (i.e., any traits that show consistent phylogenetic signal or antisignal), the application of BDPR generates phylogenetic placements of taxa that differ, sometimes greatly, from the most likely placement of those taxa if their trait information was taken into account. For example, consider a scenario where a continuous character is highly conserved across a phylogenetic tree (e.g., Pagel's $\lambda = 1$). If a pseudoposterior distribution of phylogenies for the group is generated, where some proportion of species have been placed using BDPR, they will frequently be placed in positions that are unlikely given the overall (true) process of trait evolution: in particular, the distribution will lead to occasional close relationships between taxa that may differ considerably in their trait values. Figure 1 illustrates an example dataset where BDPR lead to biased evolutionary rate estimates for a discrete character. Similar effects will be observed for phylogenetically overdispersed characters. Consider a clade where all speciation involves character displacement, such that — with complete phylogenetic knowledge — we would observe large contrasts in trait values for closely related species. Because the BDPR method places taxa without considering the most likely model of trait evolution given the observed data, we may find that many BDPR-placed taxa would show close relationships with taxa having similar trait values. These effects are generally relevant to both continuous and discrete character data.

This consideration leads to a number of predictions for patterns of trait evolution on phylogenies created with BDPR placements of unsampled taxa. If traits are conserved, BDPR phylogenies will be characterized by phylogenetically overdispersed trait values. For traits evolving under simple models of phenotypic evolution (e.g., Brownian motion for continuous traits, or a symmetric Markov model for discrete characters), BDPR placement should lead to (1) overestimation of the rate of trait evolution, (2) a reduction in phylogenetic signal of traits, (3) an acceleration in transition rates toward the tips of the tree, and (4) inflation of standardized phylogenetic contrasts for nodes near the tips of the tree.

To demonstrate the effects of BDPR on patterns of phenotypic evolution, I performed several simulation analyses using the pseudoposterior distribution of avian phylogenies from Jetz et al. (2012). Technically, this is a pseudoposterior distribution due to the fact that trees were assembled hierarchically from multiple distinct posterior distributions. This set includes 10,000 fully resolved phylogenies for all birds that effectively accounts for phylogenetic uncertainty at two levels. First, the distribution includes subtrees sampled at random from posterior distributions of phylogenies that were generated from DNA sequence data. Second, species lacking genetic data ($N = 3323 / 9993$, 33.2%) were added using the BDPR method described above ("Stage 2" in the Jetz et al. [2012] analysis).

I chose two clades of birds that differed in the percentage of taxa that were placed using the BDPR method: the Meliphagidae (honeyeaters), with 51.6% BDPR placement (90 of 178 species lack genetic data), and the Parulidae (New World wood-warblers), with 8.4% BDPR placement (10 of 119 species lack genetic data). For each clade, I randomly chose a single phylogeny from the full Jetz et al. (2012) tree distribution to represent the "true" phylogeny. I simulated trait data on these two model phylogenies using discrete and continuous models of character evolution. I then fitted evolutionary models to the trait dataset using a trial set of 1,000 fully resolved (e.g., including BDPR-placed taxa) phylogenies. Finally, I dropped all species from the trial set that had been placed using BDPR, leaving 1,000 phylogenies of taxa that had been placed with at least some genetic data. I repeated the analyses of trait evolution on this set of trees, which I refer to as "genes-only" phylogenies.

The first level of analysis, using the true tree, tells us how good our inferences would be if we knew the true generating phylogeny without error. The second level, using full BDPR phylogenies, tells us how our inferences are influenced by the use of trees with at least some BDPR taxon placements. The third level of analysis, after dropping BDPR taxa from this same set of trees, tells us how our inferences are affected by accounting for phylogenetic uncertainty in DNA sequenced-based taxon placements. I simulated 1,000 trait datasets under a Brownian motion process with a Brownian rate parameter ($\beta$) of 1.0 and a root state of 0.0 on the "true" trees for both clades. I paired each of these trait datasets with a BDPR phylogeny from the Jetz et al. (2012) distribution. I also simulated 1,000 discrete character datasets under a symmetric Markov model, with a transition rate

(q) between states sampled from a uniform distribution bounded at 0 and 0.10. To fit a model of Brownian motion to each continuous trait dataset, I computed the maximum likelihood estimate of β using the analytical equations from Garland and Ives (2000). I fitted a model of discrete character evolution using code modified from the Diversitree package for R (FitzJohn 2012). I also estimated Pagel's λ (Pagel 1997) for each dataset, a measure of the extent of phylogenetic signal in the data. I tested whether a model with λ < 1 was significantly better than the true model (λ = 1) using likelihood ratio tests.

The use of BDPR for placing unsampled taxa systematically biases inferences about evolutionary rates and phylogenetic signal (Fig. 2). For the Meliphagidae (Fig. 2, upper row), estimates of the Brownian rate parameter β are accurate when the true phylogeny is used for inference. However, with BDPR phylogenies, rates are strongly biased upwards. The median and mean values across 1000 BDPR phylogenies were 4.4 and 6.52. Fully 10.4% of all BDPR phylogenies were characterized by β estimates that exceeded 1000% of the true value. When the data are analyzed with the corresponding genes-only phylogenies, the median value of β is 0.992, and the 0.05 and 0.95 quantiles of the distribution of estimates were 0.70 and 1.47. This result indicates that the bias in evolutionary rates is almost exclusively driven by the presence of BDPR-placed taxa and not by phylogenetic uncertainty per se. Estimated values of λ are substantially and negatively displaced for BDPR phylogenies, but not for the genes-only phylogenies. This bias in λ is associated with strong statistical support for λ < 1: 99.3% of BDPR phylogenies rejected the true value (λ = 1) at the α = 0.001 level (Fig. 2E). In contrast, only 0.23% of the genes-only phylogenies showed this bias. Genes-only phylogenies are weakly biased away from the true value of λ, but this bias is trivial when compared to results from BDPR phylogenies.

Results for the Parulidae are generally similar to those for the Meliphagidae (Fig. 2, lower row). However, the overall magnitude of bias is lower and a smaller proportion of phylogenies rejected the true value of λ (note change of y-axis scale). Nonetheless, median estimates of β were approximately 150% of the generating value for BDPR phylogenies. The variance of estimates from genes-only phylogenies was greater than when the true phylogeny was used for inference, but estimates performed much more poorly when BDPR taxa were included. The true value of λ was rejected for a majority of BDPR phylogenies at the $\alpha = 0.001$ level, despite the fact that only 8% of species in these phylogenies were placed using BDPR.

Results for discrete characters are generally similar to those reported for continuous characters. I converted the estimated transition rate for each phylogeny into a ratio ($q_{est}$ / $q_{true}$) and plotted them on a log-scale, thus showing the distribution of proportional error estimates for each class of analyses (Fig. 3). For the Meliphagidae (Fig. 3A), estimated transition rates are substantially biased upwards, with a median estimated rate that was 5.3 times greater than the true rate. This distribution was highly asymmetrical, and the upper bound on proportional error estimates was very high, with 24.2% of all phylogenies showing q estimates that were at least 100 times greater than the true value. The median estimates obtained using the true and genes-only phylogenies were 1.02 and 1.05 times greater than the true value, respectively. The parulid dataset (Fig. 3B) showed a relatively weak effect of BDPR taxon placement on the distribution of q estimates.

The use of BDPR phylogenies may be conservative for diversification analyses, but their use for applications involving character data is susceptible to extreme bias (Fig. 2). I have shown, using a real distribution of BDPR phylogenies that has been used for downstream comparative analyses, that evolutionary rate parameters are consistently biased upwards and estimates of phylogenetic signal are biased downwards. The bias is marginal for discrete characters when most (92%) taxa are placed with genetic data (Fig. 3B), but it is still appreciable for continuous characters (Fig. 2B). Parameters estimated for phylogenies with relatively high BDPR placement can be inaccurate by an order of magnitude or more. For example, the maximum estimate of the Brownian rate parameter for a BDPR phylogeny was 288 times greater than the generating value ($\beta = 1$); the corresponding maximum observed values for the true and genes-only phylogenies were merely 1.32 and 2.78 times higher than the true value.

I have not specifically investigated the consequences of these biases for PGLS, trait-dependent diversification, or other purposes, but we should expect the effects to be severe for some types of analyses. All (trait) comparative methods incorporate evolutionary rates and/or covariances on some level, and for conserved traits we expect BDPR to generate inflated estimates of rate parameters as well as negatively-biased phylogenetic covariances among taxa. This latter issue is of concern for phylogenetic regression analyses, because BDPR will make comparative data appear more independent than they actually are, thus increasing the propensity for Type I error. The levels of BDPR taxon placement in datasets that have been used for comparative analysis are well within the scope of the parameter space considered here: for birds, the total fraction of taxa placed with BDPR is intermediate between the two values considered here (~33% placement; Jetz et al. [2012]). For mammals, the value is close to 50% (Fritz et al. 2009; Kuhn et al. 2011) suggesting that the extreme biases observed for the Meliphagidae are not unreasonable for mammals considered as a whole.

It is tempting to view fully-resolved BDPR phylogenies as a useful summary of phylogenetic uncertainty that can be incorporated into a range of downstream comparative analyses. However, my results demonstrate that the uncritical use of such phylogenies can lead to severe biases in the estimation of evolutionary rate parameters. If true distributions of traits show phylogenetic conservatism or overdispersion, then random polytomy resolution will alter the natural phylogenetic structure of trait data and lead to predictably biased inference. This issue is irrelevant if there is no phylogenetic structure in the traits, although if the traits lack phylogenetic structure, there would generally be no reason to worry about the phylogeny in the first place. I have shown that after pruning out BDPR-placed taxa, use of genes-only phylogenies does not appreciably bias inference, at least for the scenarios considered here. It is possible that some sampling designs using BDPR phylogenies will have relatively little impact on analyses (e.g., phylogenetic regression analyses using very sparse and phylogenetically broad taxon sets). However, the robustness of results obtained with BDPR phylogenies must be considered very carefully to ensure that strong directional effects of BDPR taxon placements have not been introduced into results.

Conclusion

We have made tremendous progress in fleshing out the deeper branches of the tree of life, but the tips of the tree remain poorly known for many groups of organisms. The majority of species-level taxa for the best-studied large clade (vertebrates) presently lack genetic data with which to estimate phylogenetic relationships. There is a danger in assuming that we can model away what we fail to understand about the seemingly inconsequential twigs on the tree of life. There is no easy solution to the present lack of data on species-level relationships, as obtaining genetic data from many of these poorly-sampled or rare taxa will require considerable time and resources. It is now possible to place taxa lacking genetic data onto phylogenies while taking character state information into account (Revell et al. 2015). Such placements will always be specific to a particular set of character state data and there will be no single posterior distribution of phylogenies that can account for phylogenetic uncertainty while being suitable for a range of downstream comparative analyses. Such approaches have utility in some contexts, particularly for diversification analyses and in the estimation of phylogenetic distinctiveness measures (Jetz et al. 2014), as both of these approaches are biased by the presence of unresolved polytomies.  However, there is ultimately no substitute for real data if we are to continue exploring the evolutionary and ecological dynamics that have generated the diversity of life on Earth.

Acknowledgements

I thank Gavin Thomas and Alison Davis Rabosky for comments on the manuscript. I declare no conflicts of interest.

Figure 1

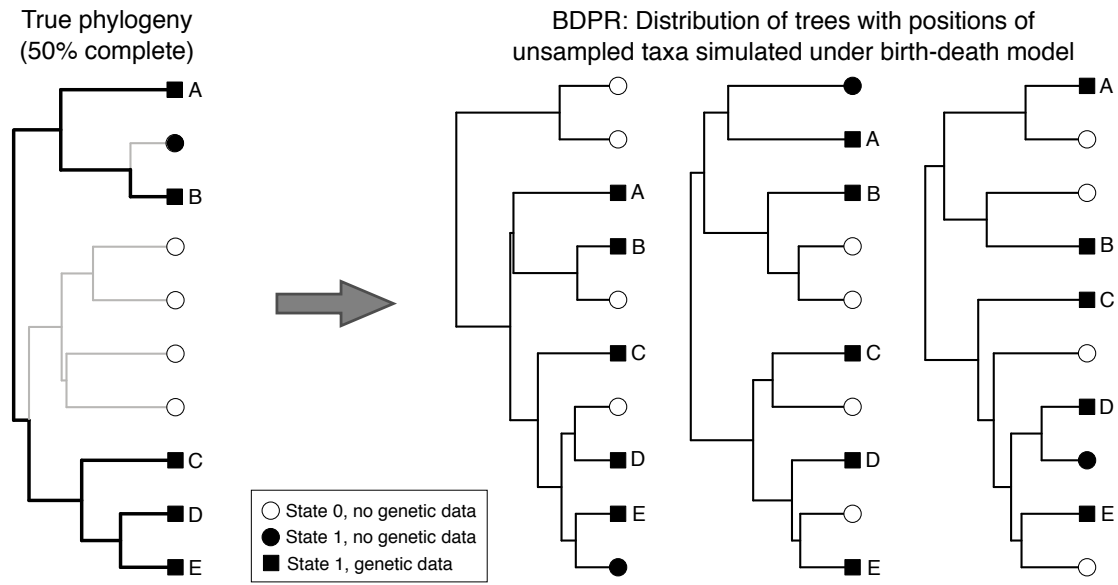

Figure 1. Birth-death polytomy resolution (BDPR) and its consequences for evolutionary rate estimation. A particular genus contains 10 species, only 5 of which have genetic data (squares; A - E). Left figure illustrates the true phylogeny, and dark lines connect taxa from the known (DNA sequence-based) phylogeny. Character states (black or white) are known for all 10 species. In this example, BDPR simulates the positions of taxa for which genetic data are lacking under a constant-rate birth-death model, while holding the DNA sequence-based topology constant. Taxa that lack genetic data (circles) are added at random to the tree without regard for character state. The true distribution of character states suggests phylogenetic conservatism and low transition rates between states, but the BDPR phylogenies show patterns of interspersed character state consistent with high transition rates. From a parsimony perspective, the BDPR phylogenies require at least three transitions to explain the character state distribution, versus one for the true phylogeny. BDPR phylogenies will be biased towards faster evolutionary rates for any characters where the true trait distribution is phylogenetically conserved.

Figure 2

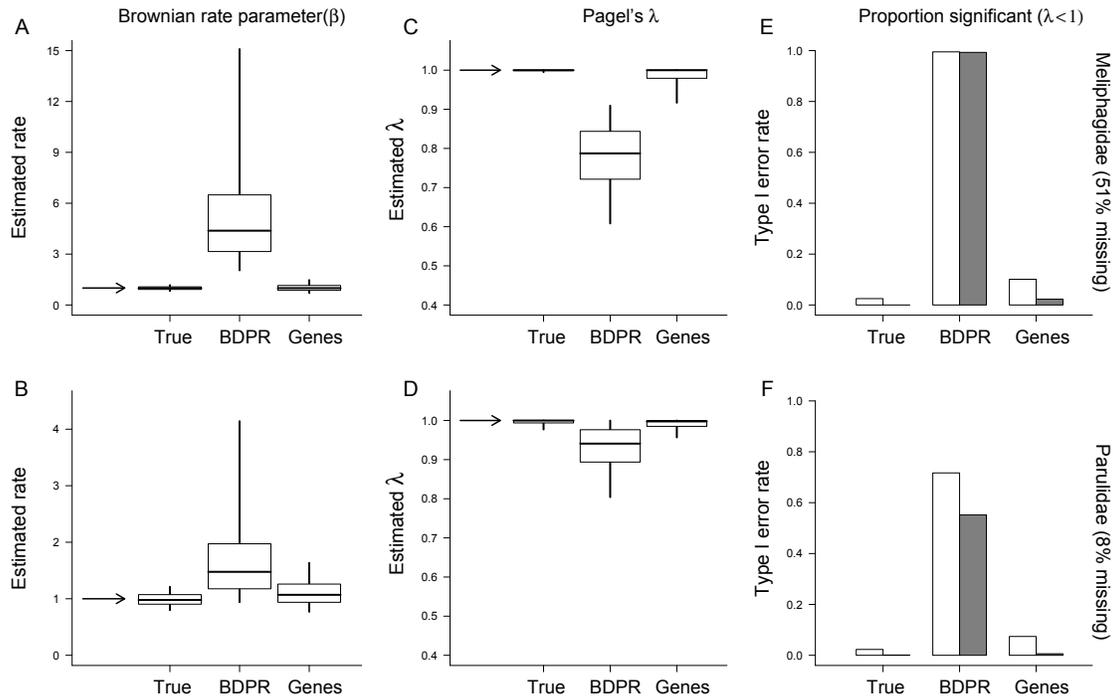

Figure 2. Estimates of evolutionary rates and phylogenetic signal for continuous characters are biased by birth-death polytomy resolution of unsampled species. Top row shows results for Meliphagidae and bottom row shows results for Parulidae. (A, B) Brownian rate parameter estimates for trait datasets analyzed with the generating phylogeny (True), phylogenies with BDPR taxon placements (BDPR), and the genes-only phylogenies (Genes). The genes-only phylogenies were created by pruning all BDPR-placed taxa from the fully resolved BDPR phylogenies and thus include phylogenetic uncertainty inherent to the DNA sequence-based placements of taxa. Whiskers denote the 0.10 and 0.90 quantiles of the distribution of estimates across 1000 phylogenies, and margins of boxes denote the corresponding 0.25 and 0.75 quantiles. Line within the box denotes the median. Arrows denote the true value of the parameter. (C, D) Estimates of phylogenetic signal ($\lambda$) in the true, BDPR, and genes-only phylogenies. (E, F) Proportion of phylogenies for which a model constrained to the true value of $\lambda$ ($\lambda = 1$) was rejected in favor of a model with $\lambda < 1$. Light and dark bars represent the proportion rejected with significance levels of 0.05 and 0.001, respectively. In all cases, phylogenies with BDPR-placements show strong positive biases in evolutionary rates and reduced phylogenetic signal relative to the generating model.

Figure 3

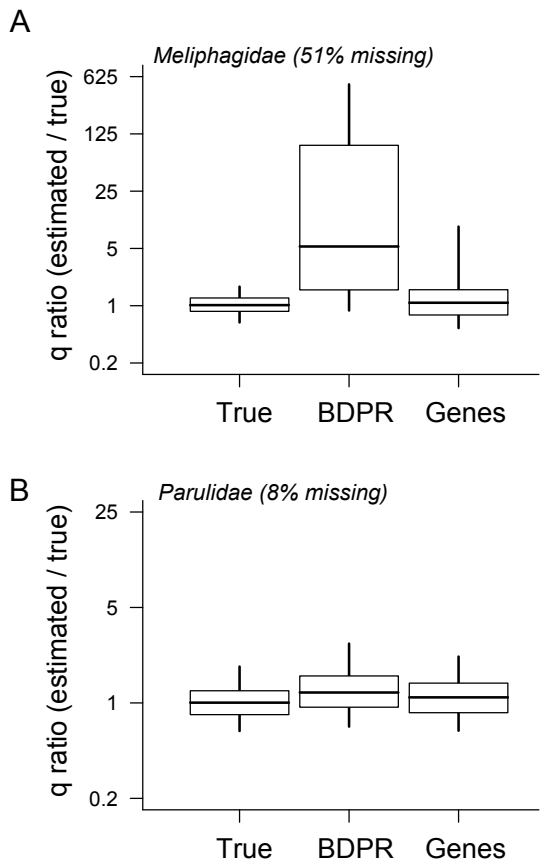

A

*Meliphagidae (51% missing)*

B

*Parulidae (8% missing)*

Figure 3. Estimates of evolutionary rates for discrete characters are biased by BDPR placement of unsampled species. (A) and (B) represent results for Meliphagidae and Parulidae, respectively. Distributions are shown as in Figure 2 and represent the ratio of estimated transition rate to true transition rate in the generating model. Due to the severity of the bias for BDPR phylogenies in the Meliphagidae (A), results are shown on a logarithmic scale. The median rate estimate for BDPR phylogenies in (A) was 5.28 time higher than the generating value, and the mean rate estimate was 103.6 times greater than the generating value. These biases are largely eliminated when the data are analyzed with the same set of phylogenies after pruning all taxa placed with BDPR (Genes).